\documentclass[aps,prl,twocolumn,showpacs]{revtex4-1}

\usepackage{graphicx}
\usepackage{color}
\usepackage{amsmath}
\usepackage{amssymb}
\usepackage[normalem]{ulem}
\usepackage[usenames,dvipsnames]{xcolor}
\usepackage{tabularx}
\usepackage{pdfpages}

\newcommand{\kB}   {k_{\rm B}}

\begin{document}

\title{Momentum-space correlations of a one-dimensional Bose gas}

\date{\today}
\author{Bess Fang$^1$, Aisling Johnson$^1$, Tommaso Roscilde$^{2,3}$,
  and Isabelle Bouchoule$^1$}
\affiliation{$^1$ Laboratoire Charles Fabry, CNRS UMR 8501, Institut
  d'Optique, Univ Paris Sud 11, 2, Avenue Augustin Fresnel, 91127
  Palaiseau, France}
\affiliation{$^2$ Laboratoire de Physique, CNRS UMR 5672, Ecole
  Normale Sup\'erieure de Lyon, Universit\'e de Lyon, 46 All\'ee
  d'€™Italie, Lyon, F-69364, France}
\affiliation{$^3$ Institut Universitaire de France, 103 boulevard
  Saint-Michel, 75005 Paris, France}

\begin{abstract}
Analyzing the noise in the momentum profiles of single realizations of
one-dimensional Bose gases, we present the  experimental
measurement of the full momentum-space density correlations $\langle
\delta n_p \delta n_{p'}\rangle$, which are related to the two-body
momentum correlation function.  Our data span the weakly interacting
region of the phase diagram, going from the the ideal Bose gas regime
to the quasicondensate regime.  We show experimentally that the
bunching phenomenon, which manifests itself as super-Poissonian local
fluctuations in momentum space, is present in all regimes.  The
quasicondensate regime is
however characterized by the presence of
negative correlations between different momenta, in contrast to
Bogolyubov theory for Bose condensates, predicting positive
correlations between opposite momenta.  Our
data are in good agreement with {\it ab-initio} calculations.
\end{abstract}


\maketitle

\paragraph*{Introduction.}
Ultracold-atom experiments have proven their efficiency as quantum simulators
of models in quantum many-body physics~\cite{Review}.  One dimensional (1D)
gases in particular are accurately simulated, as shown by the excellent
agreement between experimental results and \emph{ab initio} theoretical
predictions~\cite{armijo_mapping_2011,jacqmin_sub-poissonian_2011,
  van_amerongen_yang-yang_2008, Paredesetal2004, kinoshita_observation_2004,
  Kinoshitaetal2005, Kinoshitaetal2006, Halleretal2010}.  Among the least
understood properties of quantum many-body systems is the out-of-equilibrium
dynamics, addressed recently by several cold-atom
experiments~\cite{Kinoshitaetal2006, cheneau_light-cone-like_2012,
  langen_local_2013, fang_quench-induced_2014}.

Correlation functions are essential tools to describe the physics of a system,
as they fundamentally characterize the different phases the system can
exhibit~\cite{altman_probing_2004}.  This is particularly true for 1D gases,
where the role of fluctuations is enhanced.  For instance, the local two-body
correlation function in real space distinguishes the ideal Bose gas (IBG)
regime (characterized by bunching) from the quasicondensate (qBEC) regime
(with the absence of bunching) from the fermionized regime (characterized by
strong antibunching)~\cite{Kinoshitaetal2005}\footnote{Those regimes have also
  been indentified in~\cite{jacqmin_sub-poissonian_2011}, where the integral
  of the two-body correlation function is investigated.}.  The two-body
correlation function in an expanding Bose gas has been measured
in~\cite{Perrin2012} and can be used for thermometry in the qBEC
regime~\cite{Imambekov2009}, while higher order correlation functions permit
to identify non thermal states~\cite{Langen2015}.  Correlation functions
are also essential to describe out-of-equilibrium dynamics. For example, the
light-cone effect has been reported on the time evolution of the correlation
functions after a sudden perturbation of the
system~\cite{langen_local_2013,cheneau_light-cone-like_2012}, and the
dynamical Casimir effect was identified by studying a two-body correlation
function in~\cite{jaskula_acoustic_2012}.  Investigating the behavior of
correlation functions is thus an important issue in quantum
  simulation. However, correlation functions, especially those of higher
orders, are in general unknown theoretically, not even at
thermal equilibrium, so that further knowledge in this domain is highly
desirable.

In this letter, we investigate for the first time the full structure of the
second-order correlation function in momentum space of a 1D Bose gas at
thermal equilibrium. The measurements rely on the statistical noise analysis
of sets of momentum profiles taken under similar experimental conditions. Our
data span the weakly interacting region of the phase diagram of 1D Bose
gases~\cite{bouchoule_interaction-induced_2007}, going from the qBEC regime to
the IBG regime.  The bunching phenomenon, which manifests itself by strong,
super-Poissonian local fluctuations in momentum space, is seen in all
regimes. The qBEC regime is however characterized by the presence of negative
correlations associating different momenta, as predicted
in~\cite{bouchoule_two-body_2012}.  This contrasts with the positive
correlations between opposite momenta expected for systems with true or quasi
long-range order~\cite{mathey_noise_2009}.  In both asymptotic regimes, our
data compare well with appropriate models, while the data in the crossover are
in good agreement with Quantum Monte Carlo (QMC) simulations.  These
comparisons involve no fitting parameters.  Finally, we propose a quantitative
criterion to characterize the crossover.

\paragraph*{Experiment.}
Using an atom-chip experiment, we realise single quasi-1D ultracold $^{87}$Rb
clouds, as described in~\cite{jacqmin_momentum_2012}.  Using evaporative
cooling, we prepare atoms in the $|F=2,m_F=2\rangle$ ground state, at thermal
equilibrium in a harmonic trap whose transverse and longitudinal oscillations
frequencies are $\omega_\perp/(2\pi) \simeq 1.9$~kHz and $\omega_z/(2\pi)
\simeq 7$~Hz respectively.  The estimated population in the transverse excited
states is at most $40\%$, such that the data are indeed close to the 1D regime
of Bose gases.  We perform thermometry by fitting the measured mean \emph{in
  situ} linear density profile $\rho(z)$ and density fluctuations to the
thermodynamic predictions of the modified Yang-Yang (MYY)
model~\cite{armijo_mapping_2011, van_amerongen_yang-yang_2008}, where the
interatomic interaction is taken into account only in the transverse ground
state, modeled by a contact term of coupling constant $g=2\hbar\omega_\perp
a$, $a=5.3$~nm being the 3D scattering length.

\begin{figure*}
\includegraphics{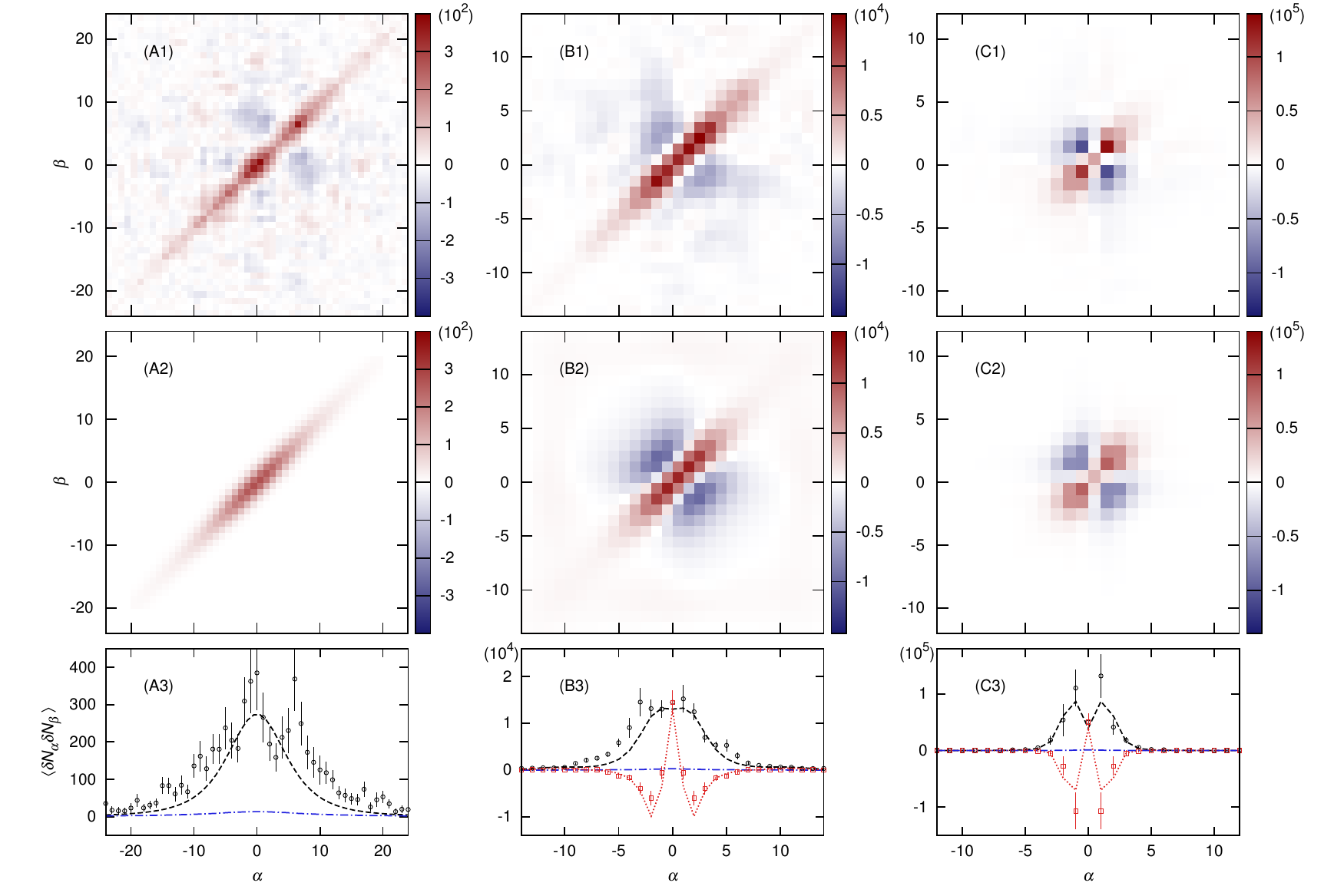}
\caption{ Momentum correlations $\langle \delta N_\alpha\delta N_\beta\rangle$
  for a gas in the IBG regime (Data A, left column), in the qBEC regime (Data
  C, right column), and in the qBEC-IBG crossover (Data B, middle column).
  The pixel size is $\Delta/\hbar= 0.15$~$\mu$m$^{-1}$.  The experimental data
  are shown in the top row.  Data A, B and C are compared with the IBG theory,
  QMC calculations, and qBEC theory respectively, at the temperature of the
  data determined by independent thermometry
  methods~\cite{SM}.  The middle row gives the computed
  momentum correlations.  The bottom row shows the diagonal cuts: the
  experimental data in circles for $\alpha = \beta$ (squares for $\alpha =
  -\beta$ for Data B and C only) are compared with their respective theory
  model in dashed (dotted) lines.  The error bars are statistical.  The
  dash-dotted lines give the shot-noise limit.  }
\label{fig.grossefigure}
\end{figure*}

A single shot of the momentum distribution $n(p)$ is obtained by imaging the
atomic cloud in the Fourier plane of a magnetic lens using the focusing
technique~\cite{shvarchuck_bose-einstein_2002,davis_yang-yang_2012,jacqmin_momentum_2012},
as detailed in the supplementary material (SM)~\cite{SM}: the spatial
distribution of the atom cloud then reflects the initial momentum
distribution~\cite{jacqmin_momentum_2012}.  These images are discretized with
a pixel size in momentum space $\Delta$.  Moreover, the resolution of the
optical system and the atomic motion during the imaging pulse are responsible
for blurring, modeled by a Gaussian impulse response function of
root-mean-square width $\delta$.  The effective atom number measured in pixel
$\alpha$ is thus $N_\alpha =\int\!dp ~n(p) {\cal A}(\alpha,p)$, where ${\cal
  A}(\alpha,p)= \int_{\Delta_\alpha}
dq~e^{-(p-q)^2/(2\delta^2)}/(\delta\sqrt{2\pi})$.  The second-order
correlation function is deduced from a set of momentum profiles taken under
similar experimental conditions.  The standard deviation of shot-to-shot
atom-number fluctuations ranges from $4\%$ at high densities to $40\%$ at low
densities. To mitigate their effect, we order profiles according to their atom
number and, for each profile, we use a running average to compute the
corresponding mean profile $\langle N_\alpha\rangle$.  Moreover, we normalize
each profile to the atom number of the running average, before computing the
fluctuations $\delta N_\alpha=N_\alpha-\langle N_\alpha\rangle$.  We finally
extract the momentum-correlation map $\langle \delta N_\alpha \delta N_\beta
\rangle$.  Fig.~\ref{fig.grossefigure} (top row) shows the results for three
different clouds lying respectively A) in the IBG regime, B) in the qBEC-IBG
crossover, and C) deep in the qBEC regime. For the data presented in this
letter, the focusing time is $\tau=25$~ms, leading to a pixel size in momentum
space $\Delta/\hbar= 0.15$~$\mu$m$^{-1}$.  The resolution is
$\delta/\Delta\simeq 1.1$~\footnote{The resolution is estimated by fitting the
  cuts of the measured correlations of Data A (IBG) in the direction where
  $p+p'=$~constant (i.e.~parallel to the antidiagonal) and the correlations
  between different pixels are introduced by the imaging resolution alone (see
  discussions on IBG).}.

\paragraph*{Ideal Bose gas regime.}
Thermometry based on \emph{in situ} density profiles indicates that Data A
lies within the IBG regime ($N=1900$, $T=109$~nK)~\cite{SM}.
Fig.~\ref{fig.grossefigure}~(A1) shows the corresponding momentum
correlations.  We observe large correlations on the diagonal $\alpha\simeq
\beta$, while $\langle \delta N_\alpha\delta N_\beta\rangle$ takes
substantially smaller and rather erratic values in the rest of the
plane~\footnote{The off-diagonal contributions may be due to the spurious
  effect induced by the large atom-number fluctuations in this low-density
  regime.}.  This is consistent with what is expected for a homogeneous IBG in
the grand canonical ensemble: since the single-particle eigenstates have well
defined momenta, the correlations between different momenta are vanishing.
Moreover, fluctuations of the occupation number $N_p$ in the state of momentum
$p$ are $\langle \delta N_p^2\rangle=\langle N_p \rangle + \langle N_p
\rangle^2 $, where the second term is the famous bunching term.  Previous
results generalise to the case of our trap clouds through a local density
approximation (LDA), as outlined in the SM~\cite{SM}, valid since the
correlation length of $\langle {\psi^\dagger}(z)\psi(z')\rangle$ is much
smaller than the cloud length $L$~\footnote{The effect of a harmonic potential
  can be computed exactly for an IBG, but we prefer to adopt the LDA, which is
  a more general approach that is applicable to interacting gases, regardless
  of the shape of the external confinement.}.  The momentum-space density
correlations is then the sum of the shot noise and bunching contributions,
\begin{eqnarray}
\langle \delta n_p \delta n_{p'} \rangle & = & \delta (p-p')\langle
n_p\rangle + B(p, p'),\\
B(p,p') & = & \left| \int dz \left\langle
\nu_{\rho(z),T}^{(h)}((p+p')/2) \right\rangle e^{i(p-p')z/\hbar}
\right |^2, \nonumber
\label{eq.B}
\end{eqnarray}
where the bunching term $B(p,p')$ uses the
momentum distribution $\nu_{\rho,T}^{(h)}(p)$ of a homogeneous gas of
temperature $T$ and linear density $\rho$, normalized to $\rho=\int
dp~\nu_{\rho,T}^{(h)}(p)$.  $B(p,p')$ takes non-zero values only for
$|p'-p|$ of the order of $\hbar/L$.   Since here $\hbar/L
  \ll \delta$, one can make the approximation $B(p,p')={\cal
    B}(p)\delta(p-p')$, where
\begin{equation}
{\cal B}(p)= 2\pi \hbar \int dz \langle \nu_{\rho(z),T}^{(h)}(p)\rangle ^2.
\label{eq.calB}
\end{equation}
Note that for a degenerate cloud, for $p$ within the width of $n(p)$, the
bunching term is much larger than the shot-noise term since $\langle
\nu_{\rho(z),T}^{(h)}(p)\rangle\gg 1$.  Finally, blurring and
discretization lead to the momentum-correlation map
\begin{equation}
\langle \delta N_\alpha \delta N_\beta\rangle = \int\!\!\int dp dp'
~{\cal A}(\alpha,p){\cal A}(\beta,p') \langle \delta n_p \delta
n_{p'}\rangle.
\label{eq.effetres}
\end{equation}

The theoretical prediction quantitatively describes our measurements, as shown
in Fig.~\ref{fig.grossefigure}~(A1-A2).  Here we evaluate Eq.~(\ref{eq.calB})
approximating $\langle \nu_{\rho(z),T}^{(h)}(p)\rangle$ by its value for
highly degenerate IBG gases: a Lorentzian of full width at half maximum (FWHM)
of $2\hbar/l_\phi$, where $l_\phi=\hbar^2\rho/(m\kB T)$.  Since correlations
between different pixels are introduced by the finite resolution
alone~\footnote{We have $\Delta \gg \hbar/L$, such that the bunching term
  introduces negligible correlations between different pixels.}, the only
relevant information is the diagonal term $\langle \delta N_\alpha^2\rangle $,
whose scaling behavior is discussed in the SM~\cite{SM}.  In
Fig.~\ref{fig.grossefigure}~(A3), we overlay the measured $\langle \delta
N_\alpha^2\rangle$ to theoretical predictions, and find a good agreement up to
statistical error of the measurement.  The fluctuations are well above the
shot-noise level, which is obtained by setting ${\cal B}(p) = 0$, showing that
this IBG is highly degenerate.

Note that the above grand-canonical analysis is legitimate since
$\hbar/l_\phi\gg \Delta \gg \hbar/L$: a pixel may be described by a subsystem
at equilibrium with the reservoir of energy and particles formed by the rest
of the cloud.

\paragraph*{Quasicondensate regime.}
The analysis of the \emph{in situ} density fluctuations~\cite{SM}, shows that
Data C lies in the qBEC regime ($N=14000$, $T \simeq 75$~nK).  The mean
density profile indicates a slightly higher temperature ($T=103$~nK), the
difference possibly coming from deviation from the Gibbs
ensemble~\cite{Langen2015,fang_quench-induced_2014}.  We show the measured
momentum correlations in Fig.~\ref{fig.grossefigure}~(C1) and its diagonal
cuts along $\alpha=\beta$ and $\alpha=-\beta$ in (C3).  We first observe that
a strong bunching in momentum space is also present here: the measured
$\langle \delta N_\alpha^2\rangle$ is well above the shot-noise level alone.
This is in stark contrast with the behavior in real space, where the qBEC
regime is characterized by the suppression of the bosonic
bunching~\cite{Esteve2006}.  Moreover, the correlation map $\langle \delta
N_\alpha \delta N_\beta\rangle$ shows strong anticorrelations around the
region $\alpha = -\beta$ (i.e.~$p'=-p$).  These features are characteristic of
the qBEC regime in a grand canonical ensemble, and have been computed for a
homogeneous gas in~\cite{bouchoule_two-body_2012}.  Since the correlation
length of the gas is much smaller than $L$~\footnote{The correlation length in
  a finite-$T$ qBEC is given by $l_\phi$ (the first-order correlation function
  decreases as $e^{-|z|/(2l_\phi)}$) and $l_\phi/L=0.035$, where $L$ is the
  FWHM of the cloud, and $l_\phi$ is evaluated at the center of the cloud.},
LDA applies and, as shown in the SM~\cite{SM}, we have
\begin{eqnarray}
\!\!\!\!\!\!\!\!\langle \delta n_p \delta n_{p'}\rangle & \!\simeq &
\delta(p-p') \langle n_{p}\rangle + B(p, p') + \langle \delta n_p
\delta n_{p'}\rangle_{\rm{reg}},
\label{eq.deltanpdeltanppgene}\\
\!\!\!\!\!\!\!\!\langle \delta n_p \delta n_{p'}\rangle_{\rm{reg}} & \!= &
\!\!\int \!dz ~\frac{l_\phi(z)^3\rho(z)^2}{(2\pi\hbar)^2} {\cal F}
\left( \frac{2l_\phi(z)p}{\hbar}, \frac{2l_\phi(z)p'}{\hbar}
\right)\!\!,
\label{eq.deltanpnppqbec}
\end{eqnarray}
where ${\cal F}$ is the dimensionless function given by Eq.~(29)
of~\cite{{bouchoule_two-body_2012}}, and $B(p,p')$ is evaluated substituting
$\nu^{(h)}_{\rho,T}(p)$ by a Lorentzian function of FWHM $\hbar/l_\phi$.  The
effect of the finite resolution and pixelization is taken into account using
Eq.~(\ref{eq.effetres}).  These predictions, plotted in
Fig.~\ref{fig.grossefigure} (C2-C3), are in quantitative agreement with
experimental data.  Note that the center-of-mass (COM) motion is decoupled
from the internal degrees of freedom in a harmonic trap, and the COM
fluctations are about twice as large as those expected at thermal equilibrium
for this data set~\footnote{The measured fluctuations might also be due to
  jittering of the kicking potential used for focusing.}. To mitigate their
effect, we post select the data by bounding the COM fluctuations.  Moreover,
since the experimental resolution is not sufficient to resolve momentum scales
of the order of $\hbar/l_\phi$, the effect of $\langle \delta n_p \delta
n_{p'}\rangle_{\rm{reg}}$ on the diagonal reduces the signal that would be
expected from bunching alone by almost a factor 10.

Our results provide the first experimental proof of the persistence of
bunching in momentum space in a qBEC, as well as the presence of negative
correlations, in particular between opposite momenta.  The latter contrasts
with the behaviour expected for a weakly interacting Bose-Einstein condensate,
where Bogoliubov theory predicts the presence of positive correlations between
opposite momenta~\cite{SM}.  The absence of opposite-$p$ positive correlations
is a clear consequence of the absence of true long range order.

The atom-number fluctuations are strongly reduced in a qBEC because of
repulsive interactions and the negative part ${\cal F}$, which concentrate on
the momentum region $p\lesssim \hbar/l_c$, enforces the reduced atom-number
fluctuations by compensating for the diagonal bunching
term~\cite{bouchoule_two-body_2012}.  In our experiment, however, one may
\emph{a priori} suspect that the measured anticorrelations could come from the
normalization procedure used in the data analysis. We rule out such a
possibility by performing several checks, detailed in the SM~\cite{SM}.  The
agreement with theory in our case is ensured by the fact that the fluctuations
$\langle \delta n_p \delta n_{p'}\rangle_{\rm{reg}}$ are dominated by the
contribution from the central part of the cloud, where $l_\phi$ is the largest
[see Eq.~(\ref{eq.deltanpnppqbec})].  It is well described by the grand
canonical ensemble as the rest of the cloud acts as a reservoir, and the
corresponding anticorrelations are much stronger than those introduced by the
normalization of the total atom number.  The negative part of the correlation
map thus reflects a local decrease of the atom-number fluctuations.

\begin{figure}
\includegraphics{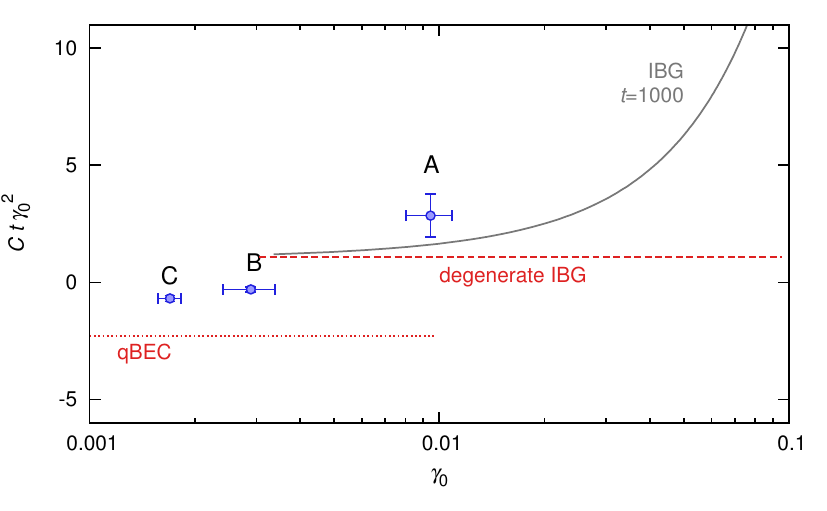}
\caption{$Ct\gamma_0^2$ versus $\gamma_0$, the interaction parameter at the
  center of the cloud.  The theoretical predictions in the limit
  $\hbar/l_\phi\gg \delta \gg \Delta$~\cite{SM} are plotted for highly
  degenerate IBG (dashed line), qBEC (dotted line) and IBG (solid line,
  computed at $t=1000$).  Circles are experimental results for Data A-C
  (corresponding to $t=1170,~1760,~920$).  The error bars account for both
  fitting (in $t$) and statistics (in $C$ and $\gamma_0$).  }
\label{fig.fonctionC}
\end{figure}

\paragraph*{In the qBEC-IBG crossover.}
While the theoretical analyses above describe reasonably well the two
asymptotic regimes of IBG and qBEC, they do not permit to investigate the
crossover in between.  To explore the crossover, we use canonical QMC
calculations~\footnote{For typical experimental parameters, the interaction
  present in our systems is not sufficiently weak for the mean-field
  predictions to be accurate~\cite{jacqmin_momentum_2012}.  We therefore
  refrain from using a classical field approach, as done
  in~\cite{bouchoule_two-body_2012}.  }.  Discretizing space allows to recast
the Lieb-Liniger model in the form of a Bose-Hubbard
model~\cite{jacqmin_momentum_2012}, which can be simulated via the Stochastic
Series Expansion with directed-loop updates~\cite{Syljuasen03}. In particular,
a double directed-loop update allows one to compute the momentum correlations
$\langle \delta n_p \delta n_{p'}\rangle$.  Blurring and pixelisation is then
applied according to Eq.~(\ref{eq.effetres}).  The features of $\langle \delta
n_p \delta n_{p'}\rangle$ are mainly washed out at the level of the
experimental resolution~\footnote{The regular part of the momentum correlation
  cancels almost completly the bunching on the diagonal.}, demanding a very
high numerical precision on $\langle \delta n_p \delta n_{p'}\rangle$ to
properly evaluate the discretised correlation map.  The results for the
parameters of Data B ($N=7000$, $T=144$~nK), shown in
Fig.~\ref{fig.grossefigure}~(B2) and (B3), reproduce quantitatively the
features seen in the experimental data, shown in
Fig.~\ref{fig.grossefigure}~(B1) and (B3).  Namely, the bunching phenomenon
remains prominent on the $\alpha=\beta$ diagonal, while the anticorrelations
along the $\alpha=-\beta$ is less pronounced than what is found for Data C.

\paragraph*{Quantifying the crossover.}
As shown in the SM~\cite{SM}, Eqs.~(\ref{eq.deltanpdeltanppgene}) and
(\ref{eq.deltanpnppqbec}) generalize the computation of the momentum
correlations to the whole parameter space, provided ${\cal F}$ is now a
function of the reduced temperature $t=2\hbar^2 k_{\rm B}T/(mg^2)$ and the
interaction parameter $\gamma (z)=mg/\big(\hbar^2\rho (z)\big)$.  
${\cal F}$ interpolates between 0 in the IBG regime ($t\gg 1$ and 
$t\gamma^{3/2}\gg 1$)
and Eq.~(29)
of~\cite{bouchoule_two-body_2012} in 
the qBEC regime ($t\gamma^{3/2}\ll
  1$ and $\gamma \ll 1$).
For the
experimental resolution of this paper, however, one cannot isolate the
contribution of ${\cal F}$ from that of the bunching term.  We thus consider
the experimental quantity
\begin{equation}
C={\sum_\alpha \langle \delta N_{\alpha}\delta N_{-\alpha}\rangle }/
{\langle N_{0}\rangle}.
\label{eq.functionC}
\end{equation}
As derived in the SM~\cite{SM}, in the limit $\delta$,$\Delta \ll
\hbar/l_\phi$ and $\delta \gg \Delta$, $C$ depends only on $t$ and $\gamma_0
\equiv \gamma (z=0)$.  For a highly degenerate IBG, we find $C\simeq
1.08/(t\gamma_0^2)$, whereas $C\simeq -2.28/(t\gamma_0^2)$ for a qBEC.  These
asymptotic behaviors are shown as dashed and dotted lines in
Fig.~\ref{fig.fonctionC}.  The solid line gives the prediction for an IBG
(where $Ct\gamma_0^2$ now depends on $t$ and $\gamma_0$) at
$t=1000$~\footnote{Note that numerical computation of $C$ is however beyond
  the precision of our QMC calculations, mainly as a consequence of the coarse
  resolution of the measurements.}.  Fig.~\ref{fig.fonctionC} also displays
the experimental values of $Ct\gamma_0^2$ for Data A-C. However, since
$\hbar/l_\phi \gg \delta \gg\Delta $ is not satisfied for our data sets, the
above theoretical predictions are not expected to quantitatively agree with
the experimental data.  Moreover, comparing different data sets is delicate
since they correspond to different values of $\delta/l_\phi$.

\emph{Outlook}. A future extension of our study of two-body correlations in
momentum space concerns the fermionized regime of 1D Bose gases, where quantum
fluctuations, difficult to observe in momentum space for weakly interacting
gases, might have measurable effects.  The study of correlations in momentum
space at thermal equilibrium could serve as a reference for the investigation
of non-thermal states and that of out-of-equilibrium dynamics arising, for
example, from a quench of the coupling constant $g$.  Correlations in momentum
space have also been proposed as a probe of Hawking-like radiation generated
by a sonic black-hole~\cite{PhysRevA.90.033607}, and the results of this paper
are certainly relevant for this quest.

\emph{Acknowledgements.} This work has been supported by Cnano IdF and by the
Austro-French FWR-ANR Project I607.  T.R. acknowledges support of the ANR
(``ArtiQ'' project). QMC simulations have been performed on the PSMN cluster
(ENS Lyon).

\onecolumngrid\newpage



\section*{Supplementary material}

# SUPPLEMENTARY MATERIAL: Two-body correlation function in momentum space of a one-dimensional Bose gas

## I. FOCUSING TECHNIQUES TO ACCESS THE MOMENTUM DISTRIBUTION OF THE CLOUD

Measuring the longitudinal momentum distribution of 1D gases is *a priori* possible by switching off all confinements and performing the so-called time-of-flight measurements. Transverse expansion occurs in a time of the order of $1/\omega_\perp$, where $\omega_\perp$ is the initial transverse confinement frequency, much shorter than the time scales of the longitudinal motion. This ensures that interactions are effectively switched off almost instantaneously with respect to the longitudinal motion. The atoms then experience a ballistic flight in the longitudinal direction. For a sufficiently long ballistic expansion time, the cloud size becomes much larger than its initial value and the density distribution is homothetic to the momentum distribution. However, reaching this far field regime requires a long time of flight, which is difficult to implement in practice. This is most stringent in the qBEC regime. Indeed, the width of the momentum distribution of a qBEC of temperature $T$ and peak linear density $\rho_0$ is of the order of $m k_B T/(\hbar \rho_0)$, while the initial size of the cloud, harmonically confined with a frequency $\omega_z/(2\pi)$, is about $\sqrt{g\rho_0/m}/\omega_z$. Thus, one finds that the far field regime is obtained only for times much larger than $\tau_{\rm ff} \simeq \sqrt{g\rho_0/m\hbar}\rho_0/(k_B T\omega_z)$. Since in the qBEC regime, $\sqrt{g\rho_0/m\hbar}\rho_0/(k_B T) \gg 1$, one has $\tau_{\rm ff} \gg 1/\omega_z$. The far field is reached only after very long times.

To circumvent this difficulty, we image the cloud in the Fourier plane of an atomic lens, using the focusing techniques [1, 2], whose implementation in our experiment is detailed in [3]. The idea is to apply a pulse of strong harmonic confinement in the longitudinal direction, of frequency $\omega_c$ and duration $\tau$, which is sufficiently short that the atomic displacement is negligeable during the pulse. This is in practice possible on our experiment by using magnetic field generated by a combination of DC and AC (modulated at 200 kHz) currents, allowing an independent control of the longitudinal and transverse potentials [4]. The effect of this harmonic pulse is then to imprint a momentum kick on the atoms $\Delta p = -m\omega_c^2 z\tau$, which is proportional to their distance $z$ from the center of the harmonic potential, and to the pulse duration $\tau$. After a time $\tau_f = 1/(\omega_c^2\tau)$ of ballistic expansion, called the focusing time, the effect of the initial spatial spread of the cloud is effectively erased, and the density distribution is homothetic to the initial momentum distribution. We use $\omega_c \simeq 2\pi \times 38$ Hz, $\tau \simeq 0.7$ ms, and $\tau_f = 25$ ms for the data presented in this article. The above classical analysis also holds at the quantum level since the quantum evolution of the Wigner function in phase space is the same as that of a classical phase space distribution as long as only quadratic potentials are involved. Note that an alternative way to measure the momentum distribution is the Bragg scattering method (see e.g. [5] and references therein). However, this method does not provide the full momentum distribution within each experimental shot and thus cannot be used to measure momentum correlations.

## II. MOMENTUM CORRELATIONS USING THE LOCAL DENSITY APPROXIMATION

Here, we outline the derivation of the key equations in the main text.

Expressing $n(p)$ in terms of the field operator, we find

$$\langle \delta n_p \delta n_{p'} \rangle = \frac{1}{(2\pi\hbar)^2} \int d^4z \; e^{ip(z_1-z_2)/\hbar} e^{ip'(z_3-z_4)/\hbar} \left( \langle \psi_1^\dagger \psi_2 \psi_3^\dagger \psi_4 \rangle - \langle \psi_1^\dagger \psi_2 \rangle \langle \psi_3^\dagger \psi_4 \rangle \right), \quad (1)$$

where $d^4z \equiv dz_1 dz_2 dz_3 dz_4$ and $\psi_i$ is a short-hand notation for $\psi(z_i)$. Let us assume the gas has a finite correlation length, that we denote $l_c$. Note that this is true for a finite temperature qBEC and a highly degenerate IBG, and $l_c$ is of the order of the phase correlation length $l_\phi = \hbar^2\rho/(m k_B T)$. The four-point correlation function can then be written as a sum of singular and regular terms [6],

$$\langle \psi_1^\dagger \psi_2 \psi_3^\dagger \psi_4 \rangle - \langle \psi_1^\dagger \psi_2 \rangle \langle \psi_3^\dagger \psi_4 \rangle = \langle \psi_1^\dagger \psi_4 \rangle \delta(z_2-z_3) \quad (2)$$
$$+ \langle \psi_1^\dagger \psi_4 \rangle \langle \psi_3^\dagger \psi_2 \rangle$$
$$+ \tilde{G}_2(1,2,3,4).$$

The first term on the right-hand side is the shot noise. The second is the bunching term, describing the exchange process due to Bose quantum statistics. In infinite uniform systems, both give rise to singular momentum correlations. The last term is what remains and gives a regular contribution to $\langle \delta n_p \delta n_{p'} \rangle$ as soon as the gas has a finite correlation length. It describes the binary elastic-scattering processes, and goes to zero whenever one of the relative coordinates is much greater than the correlation length. Isolating the contribution of each of these three terms in Eq. (1), we write

$$\langle \delta n_p \delta n_{p'} \rangle = S(p, p') + B(p, p') + \langle \delta n_p \delta n_{p'} \rangle_{\rm reg}. \quad (3)$$

The contribution of the shot-noise term $S(p, p')$ is

$$S(p, p') = \delta(p - p') \langle n_p \rangle. \quad (4)$$

To compute the contribution of the bunching term $B(p, p')$, we use a local density approximation (LDA),

which relies on the fact that the correlation length is much smaller than the cloud size $L$. We then find that

$$B(p,p') = \left| \int dz \left\langle \nu^{(h)}_{\rho(z),T}\big((p+p')/2\big) \right\rangle e^{i(p-p')z/\hbar} \right|^2. \quad (5)$$

Here, $\nu^{(h)}_{\rho,T}(p)$ is the momentum distribution of a homogeneous gas of linear density $\rho$, normalized to $\int dp\ \nu^{(h)}_{\rho,T}(p) = \rho$. For a given $p$, $\langle \nu^{(h)}_{\rho(z),T}\big((p+p')/2\big)\rangle$ has an extent in $z$ on the order of $L$. $B(p,p')$ thus has a width in $p-p'$ on the order of $\hbar/L$. Since the typical size of our 1D gases is $L \sim 100\ \mu$m, and our resolution in momentum space $\delta$ is $\delta/\hbar \simeq 0.15\ \mu\text{m}^{-1} \gg 1/L$, one can make the approximation

$$B(p,p') \simeq \delta(p-p')\mathcal{B}(p), \quad (6)$$

where $\mathcal{B}(p) = 2\pi\hbar \int dz\ \langle \nu^{(h)}_{\rho(z),T}(p)\rangle^2$ is also given in the main text.

It is instructive at this stage to look at the scaling of the experimental signal predicted by the above results. The effect of discretization and resolution is taken into account through Eq. (3) of the main text. Within the approximation of Eq. (6), all information lies in the diagonal, which writes

$$\langle \delta N_\alpha^2 \rangle \simeq \int dp\ \mathcal{A}^2(\alpha,p)\big(\langle n_p\rangle + \mathcal{B}(p)\big). \quad (7)$$

For an infinite resolution, i.e. $\delta \to 0$, we find that the bunching term is on the order of $\langle \delta N_\alpha^2 \rangle \simeq \Delta L (\rho l_c)^2$. We recover here the prediction from a semi-classical analysis for an IBG: the typical occupation per mode is $n \simeq Nl_c/L = \rho l_c$, and the number of modes contributing to a pixel is $M \simeq \Delta L/(2\pi)$, such that the bunching phenomenon should produce atom number fluctuations $\langle \delta N_\alpha^2 \rangle \simeq Mn^2 \simeq \Delta L(\rho l_c)^2 \simeq \rho l_c \langle N_\alpha\rangle$. We thus find that as soon as the gas is highly degenerate, i.e. for $\rho l_c \gg 1$, the bunching term is much larger than the shot-noise term.

To compute the contribution of the regular term, we apply LDA again. Together with dimensional analysis, we have

$$\langle \delta n_p \delta n_{p'} \rangle_{\text{reg}} = \int dz\ \frac{l_\phi^3 \rho^2}{(2\pi\hbar)^2} \mathcal{F}\left(\frac{2l_\phi p}{\hbar}, \frac{2l_\phi p'}{\hbar}; t, \gamma\right), \quad (8)$$

where $l_\phi \equiv l_\phi(z) = \hbar^2\rho(z)/(mk_B T)$, $\mathcal{F}$ is a dimensionless function, $t$ is the reduced temperature parameter, and $\gamma \equiv \gamma(z) = mg/(\hbar^2\rho(z))$ is the local interaction parameter. In the IBG regime (i.e. for $t\gamma^{3/2} \gg 1$ and $t \gg 1$), Wick's theorem ensures that $\widetilde{G}_2$ vanishes, such that

$$\mathcal{F}(q,q') = 0. \quad (9)$$

In the qBEC regime (i.e. for $t\gamma^{3/2} \ll 1$ and $\gamma \ll 1$), the results obtained for homogeneous gases in [6] show that

$$\mathcal{F}(q,q') = \frac{256}{(q^2+1)^2(q'^2+1)^2[(q+q')^2+16]} \quad (10)$$
$$\times \big[(q^2+3qq'+q'^2)qq' - 2(q^2-qq'+q'^2) - 7\big].$$

Note that in [6] the crossover between the IBG and the qBEC regimes is investigated by treating $\psi$ as a classical field, in which case $\mathcal{F}$ reduces to a function of three parameters only: $\mathcal{F}(q,q';t,\gamma) = \tilde{\mathcal{F}}(q,q',\sqrt[3]{4/(t^2\gamma^3)})$.

## III. IBG TO QBEC CROSSOVER: $C$ FUNCTION

To investigate the behavior of the function $C$ defined by Eq. (6) of main text, let us first assume that $\hbar/L$ is much smaller than all other momentum scales. The momentum correlations can thus be written as

$$\langle \delta n_p \delta n_{p'}\rangle = \delta(p-p')\big(\langle n_p\rangle + \mathcal{B}(p)\big) + \langle \delta n_p \delta n_{p'}\rangle_{\text{reg}}, \quad (11)$$

where $\langle \delta n_p \delta n_{p'}\rangle_{\text{reg}}$ is given by Eq. (8). Separating the contribution of both the shot-noise and the bunching terms from that of the regular term, we can write $U = C\langle N_0\rangle \equiv U_i + U_r$, where

$$U_i = \int dp\ \mathcal{R}(p,p)\big(\langle n_p\rangle + \mathcal{B}(p)\big), \quad (12)$$

$$U_r = \iint dpdp'\ \mathcal{R}(p,p')\langle \delta n_p\delta n_{p'}\rangle_{\text{reg}}, \quad (13)$$

and

$$\mathcal{R}(p,p') = \sum_\alpha \mathcal{A}(\alpha,p)\mathcal{A}(-\alpha,p'). \quad (14)$$

For a given value of $p'+p$, the function $\mathcal{R}(p,p')$ is periodic in $p$: $\mathcal{R}(p+\Delta, p'-\Delta) = \mathcal{R}(p,p')$. Its width in $p'+p$, depending on the distance of $p$ to a pixel center, is typically on the order of $\max(\delta,\Delta)$.

The following asymptotic expressions can be deduced. In the limit of $\Delta, \delta \ll \hbar/l_c$, we have

$$U_r \simeq \Delta \int dp\ \langle \delta n_p\delta n_{-p}\rangle_{\text{reg}}, \quad (15)$$

regardless of the resolution (i.e. of the ratio $\delta/\Delta$). In the same limit, we also have

$$U_i \simeq \Delta(\langle n_0\rangle + \mathcal{B}(0))\mathcal{I}, \quad (16)$$

where $\mathcal{I} = \int dp\ \mathcal{R}(p,p)/\Delta$. The value of $\mathcal{I}$ depends on $\delta/\Delta$, and the asymptotic behavior is given by

$$\begin{cases} \mathcal{I} = 1 \text{ for } \delta/\Delta \ll 1, \\ \mathcal{I} \simeq \frac{1}{2} \text{ for } \delta/\Delta \gg 1. \end{cases} \quad (17)$$

Finally, for $\hbar/l_c \gg \delta \gg \Delta$, one obtains

$$C = \frac{1+\mathcal{B}(0)/\langle n_0\rangle}{2} + \int dp\ \langle \delta n_p\delta n_{-p}\rangle_{\text{reg}}/\langle n_0\rangle. \quad (18)$$

Using LDA and transforming integrals over $z$ into integrals over the chemical potential $\mu$, we find that $C$ is an intensive quantity which only depends on $t$ and $\gamma_0 \equiv \gamma(z=0)$. For a highly degenerate IBG, using the equation of state $\rho \simeq k_B T\sqrt{m}/(\hbar\sqrt{2|\mu|})$ and $\nu^{(h)}_{\rho,T}(0) \simeq \hbar\rho^2/(\pi mk_B T)$, we find that $C \simeq 0.54\hbar^2\rho^2/(mk_B T)$. For a qBEC, using $\rho \simeq \mu/g$ and $\nu^{(h)}_{\rho,T}(0) \simeq 2\hbar\rho^2/(\pi mk_B T)$, we find $C \simeq -1.14\hbar^2\rho^2/(mk_B T)$.

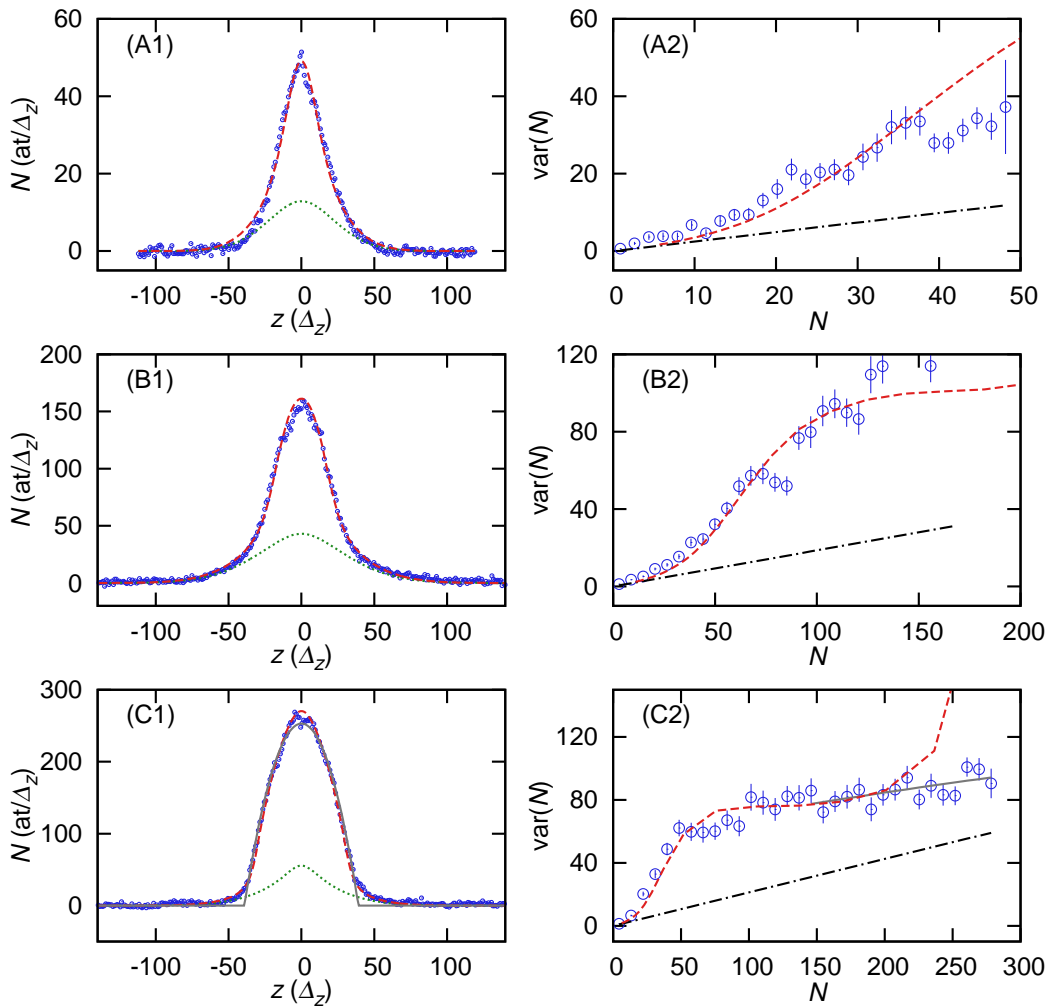

FIG. 1. Thermometry of data sets (from top to bottom, Data A, B and C). Left: density profiles. $\Delta_z = 2.7$ $\mu$m is the pixel size in real space. Fitting the measured mean profile (circles) with the MYY theory (dashed lines) yields $T = 109$ nK for Data A, $T = 144$ nK for Data B and $T = 103$ nK for Data C. The dotted lines are the contribution of the transverse excited states. We also include the zero-temperature Thomas-Fermi profile (solid line) for Data C. Right: atom-number fluctuations. Experimental data (circles) are compared to the prediction from MYY equation of state at the temperature deduced from the profile. The dash-dotted lines show the shot-noise limit. For Data C, the fit to the qBEC predictions (solid line) gives $T = 75$ nK.

## IV. CHARACTERIZATION OF THE DATA

The data are characterized using a statistical ensemble of about 100 *in-situ* images taken in the same experimental conditions. Temperature is extracted by analysing either the mean density profile or the density fluctuations. With our resolution in the atom-number calibration and the fluctuation measurements, we estimate the accuracy of our temperature measurement to be of about 20% [7].

*Data A and B.* The mean *in situ* density profiles. for Data A and B ($N = 1900$ and $N = 7000$) are shown in Fig. 1 (A1) and (B1) respectively. We fit our measurements (circles) to the MYY theory (dashed lines), and obtain $T = 109$ nK and $T = 144$ nK. The density fluctuations, shown in Fig. 1 (A2) and (B2), are in agreement

with the thermodynamic predictions using the MYY theory at the temperature obtained from the profile. For Data A, the density fluctuations at large densities do not saturate, which is characteristic of the IBG regime. On the other hand, the measured fluctuations are well above shot-noise level (dash-dotted line), indicating that the gas is highly degenerate. Data B shows the onset of saturation of the density fluctuations at large densities, indicating that this data set lies within the IBG-qBEC crossover.

*Data C.* The mean *in situ* density profile of this data set ($N = 14000$) is shown in Fig. 1 (C1). We fit our measurements (circles) to the MYY theory (dashed lines), and obtain $T = 103$ nK. However, the central part of the profile follows very well the zero-temperature

Thomas-Fermi (TF) profile (grey solid line, nearly overlapping with the MYY profile). The information about the temperature lies in the (small) wings, where the MYY profile differs from the TF profile. The uncertainty on the temperature obtained from a fit of such a density profile is large, and more precise information can be extracted from density fluctuations, shown in Fig. 1 (C2). The MYY predictions in the qBEC regime suffer from two weaknesses at large linear density. First, this model does not account for the inflation of the transverse wave function of the transverse ground state, which becomes noticeable when $\rho a$ approaches unity. Second, the treatment of the first transverse excited states as an IBG no longer captures the real behavior when $\rho$ becomes sizable, with the predicted density even diverging as $\rho a$ approaches $1/2$. For these reasons, it is preferable to consider the qBEC equation of state $\mu = \hbar\omega_\perp \left(\sqrt{1 + 4\rho a} - 1\right)$ instead of that of the MYY model in qBEC regime at large densities. The fit (solid line) of the measured density fluctuations gives $T = 75$ nK. Note that the failure of the MYY model is first visible on the predicted fluctuations [see the divergence of the dashed line in Fig. 1 (C2)], proportional to the derivative of the equation of state, before the predicted density profile is affected.

## V. EFFECT OF NORMALIZATION

The experimental realizations of 1D Bose gases as single systems do not correspond to the grand-canonical-ensemble description: the atomic cloud is not in contact with a reservoir of energy or particle at a well defined temperature or chemical potential. The shot-to-shot fluctuations of the total atom number are linked to the preparation procedures. We remove such fluctuations by normalizing each experimental profile to the total atom number of the corresponding mean profile. This procedure would be exact if the momentum distribution satisfied a scaling form of the kind $n(p; N_1)/N_1 \approx n(p; N_2)/N_2$. However, this is not the case in degenerate 1D Bose gas, since the width of the momentum distribution is proportional to $1/\rho$. One then expects that the normalisation procedure induces extra-fluctuations whose variance is proportional to the variance of atom number within the set of data used to compute the averaged profile. Note that one also expects that this normalisation procedure introduces negative correlations between small $p$ and large $p'$ components. This may partly explain the difference between theoretical and experimental data in Fig. 1 of the main text.

Moreover, the normalization procedure ensures a vanishing sum of the momentum correlation map $\langle \delta N_\alpha \delta N_\beta \rangle$, and hence it introduces additional negative regions. Since the area covered by such negative regions in the plane $(\alpha, \beta)$ scales typically as $N_\Delta^2$ (where $N_\Delta$ is the width of the momentum distribution in units of pixels), their amplitude is negligible as long as $N_\Delta \gg 1$. While the

condition $N_\Delta \gg 1$ is well fulfilled for the Data A, one may worry about the effect of normalisation for the Data C. In this appendix, we show that the negative correlations observed in Data C cannot be accounted for solely by the effect of normalization.

We first compute the anticorrelations that would arise due to the normalization procedure for a gas confined in a box, showing that it would be difficult to differentiate between the effect of normalization and the expected anticorrelations within the grand-canonical-ensemble description of a homogeneous qBEC. The experimental realization of the qBEC is however inhomogeneous, and in the second part of this appendix, we make several numerical checks to confirm that the normalization procedure does not account for the negative correlations observed in the data. In fact, the observed anticorrelations arise mainly due to the central part of the cloud, where $l_\phi$ is the largest.

### A. Effect of normalization on a homogeneous cloud

Let us consider a cloud confined in a box potential of length $L$, whose momentum correlations are assumed to be purely described by the bosonic bunching. For simplicity, we will disregard the effect of resolution, assuming $\delta = 0$. The number of momentum states contributing to a pixel of size $\Delta$ is $M = \Delta L/(2\pi)$. Thus, for $\Delta \gg \hbar/L$, the fluctuations $\delta N_\alpha$ of the atom number in a pixel, which fulfill

$$\langle \delta N_\alpha^2 \rangle = \langle N_\alpha \rangle^2/M, \tag{19}$$

are small compared to $\langle N_\alpha \rangle^2$. Linearizing in $\delta N_\alpha$, the normalization procedure to a total atom number $N$, changes the momentum profile to $\widetilde{N}_\alpha = \langle N_\alpha \rangle + \delta N_\alpha - \frac{\langle N_\alpha \rangle}{N} \sum_\beta \delta N_\beta$, leading to

$$\begin{aligned} \langle \delta \widetilde{N}_\alpha \delta \widetilde{N}_\beta \rangle =\ & \langle \delta N_\alpha^2 \rangle \delta_{\alpha,\beta} - \frac{\langle N_\alpha \rangle}{N} \langle \delta N_\alpha^2 \rangle - \frac{\langle N_\beta \rangle}{N} \langle \delta N_\beta^2 \rangle \\ & + \frac{\langle N_\alpha \rangle \langle N_\beta \rangle}{N} \sum_\gamma \langle \delta N_\gamma^2 \rangle \end{aligned} \tag{20}$$

Replacing $\langle N_\alpha \rangle$ by a Lorentzian with a FWHM of $2\hbar/l_\phi$ in Eq. (19) and (20), we obtain $\langle \delta \widetilde{N}_\alpha \delta \widetilde{N}_\beta \rangle = \langle \delta N_\alpha^2 \rangle \delta_{\alpha,\beta} + (\Delta L/(2\pi))^2 \mathcal{F}_n(2l_\phi p/\hbar, 2l_\phi p'/\hbar)(\rho l_c)^3/N$ where

$$\mathcal{F}_n(q, q') = \frac{-64(1 + q^2) - 64(1 + q'^2) + 16(1 + q'^2)(1 + q^2)}{(1 + q^2)^2(1 + q'^2)^2}. \tag{21}$$

This function is plotted in Fig. 2 (a). For a homogeneous qBEC in a grand canonical ensemble, one expects similar fluctuations, but the function $\mathcal{F}_n$ should be replaced by the function $\mathcal{F}$ of Eq. (10), plotted in Fig. 2 (b). These two functions have the same integral in the $(q, q')$-plane, and their value at $q = q' = 0$ is identical. Although they differ slightly in shape ($\mathcal{F}_n$ is even in both $q$ and $q'$ whereas $\mathcal{F}$ is not), it would be difficult to distinguish between them in practice.

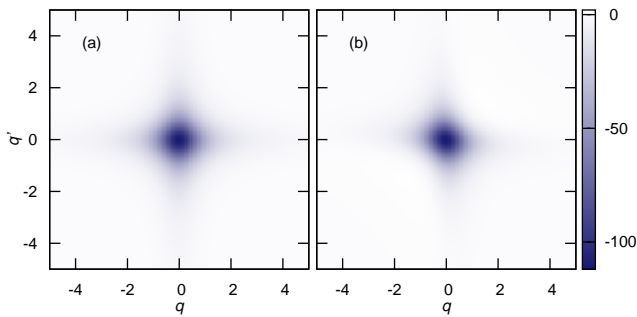

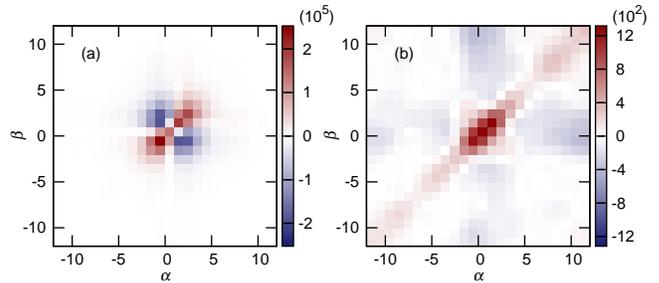

FIG. 2. (a) function $\mathcal{F}_n$ [see Eq. (21)] that describes the effect of normalization to a given total atom number, for a homogeneous gas that exhibits only the bosonic bunching in its momentum-space fluctuations. $\mathcal{F}_n$ is even in both $q$ and $q'$, so that it has a symmetry upon reflection about the lines $q = 0$ and $q' = 0$. (b) function $\mathcal{F}$ [see Eq. (10)] that describes the momentum-space anticorrelations of a homogeneous qBEC in grand canonical ensemble. These two functions have the same integral in the $(q, q')$-plane, and the same value at the origin, and would be difficult to distinguish in practice.

FIG. 3. Effect of normalizing the data. (a) Data C analyzed without normalization procedure. The total atom number of the experimental data fluctuates with a standard deviation of $\sim 4\%$. (b) Momentum correlations for Data A, rebinned and blurred in order to have similar value of $\Delta l_c$ and $\delta l_c$ as Data C. The color scale of these two plots are symmetric so that that the zero-crossing occurs at the same color. The absence of the dark blue regions in (b) indicates that the negative correlations introduced by normalization applied to Data A would not explain the negative correlations observed in Data C.

## B. Effect of normalization on a inhomogeneous cloud

Here we focus on Data C, and present different analyses showing that the negative regions observed are not accounted for by the normalization of the data.

Firstly, we numerically simulate the effect of normalization, assuming a statistical ensemble that would exhibit a pure bunching. The negative correlations induced by the normalization are lower than those measured by about a factor of 5.

Secondly, the total atom number of the experimental data fluctuates with a standard deviation of $\sim 4\%$. Without normalizing, the momentum correlations are shown in Fig. 3 (a), which has similar features as Fig. 1 (C1) of the main text but differ slightly in the extreme values.

In addition, we see that the normalization procedure applied to Data A does not introduce anticorrelations as large as those observed in Data C. Since Data A lies within the IBG regime, the anticorrelations are mostly a consequence of the normalization, whose effect depends on the values of $\Delta l_c$ and $\delta l_c$, where $\hbar/l_c$ is the width of the momentum distribution. In order to compare Data A and C on equal footing, we blur and rebin Data A (which has a broader momentum distribution) such that its values of $\Delta l_c$ and $\delta l_c$ are similar to those of Data C, and show the resulting momentum correlations in Fig. 3 (b). We observe that the anticorrelations are much weaker than the positive correlations in rebinned Data A than those in Data C, seen as the absence of dark blue in Fig. 3 (b). This further eliminates the possibility that the normalization is responsible for the observed anticorrelations in Data C.

We remark that a similar procedure could in principle be employed to compute numerically (with QMC calcu-

lations) the value of the quantity $C$ in order to compare across different data sets on an equal footing. However, the smaller $l_c$ value of Data A and the convergence of function $C$ requires to access the momentum distribution at larger momenta and at a higher signal-to-noise ratio. Both items are beyond the current data sets and will be reserved for future investigations.

## VI. PREDICTION FOR TRUE OR QUASI LONG RANGE ORDER

In [8], Mathey and colleagues derived the noise correlations $G(k, k')$, where $k = p/\hbar$, assuming true long range order (TLRO) : expanding the field operator up to the second order around its mean value, $G(k, k')$ is computed with a Bogoliubov treatment. $G(k, k')$ is then the sum of two singular terms, the diagonal in $\delta(k - k')$ and the anti-diagonal in $\delta(k + k')$, each of them being positive. The prediction of a positive antidiagonal can be understood, at zero-temperature, as the result of the elementary process of condensate depletion, namely the momentum-conserving production of opposite-momentm pairs leaving the $k = 0$ condensate. We applied this formula to a homogeneous one-dimensional Bose gas with a temperature and linear density equal to those found at the centre of the cloud of Data C. The resulting correlation map, after convolving with our imaging response function, is shown in Fig. 4. The result is approximately symmetric around the $k = 0$ and $k' = 0$ lines, which is expected deep in the phononic regime. In particular, the anti-diagonal is about the same as the diagonal.

The Hohenberg-Mermin-Wagner theorem prevents

TLRO in a 1D gas at thermodynamic limit. At zero temperature however, the correlation length $l_\phi$ of the first order correlation function $g^{(1)}$ diverges, $g^{(1)}$ decreasing only algebraically. In such a case the results obtained for TLRO are still qualitatively correct, as shown by the Luttinger Liquid calculations presented in [8], although the singularities around the anti-diagonal and the lines $k = 0$ and $k' = 0$ get broadened.

For our experimental data, the temperature is sufficiently high so that $l_\phi$ is much smaller than the size of the cloud, such that the condition for the above calculations is not satisfied. Indeed, Fig. 4 differs strongly from our experimental data. The most salient difference lies in the symmetry of the correlation maps : whereas Fig. 4 is symmetric around the lines $k = 0$ and $k' = 0$, the anti-diagonal and the diagonal being equal, the experimental correlation map shown in Fig. 1 (C1) of the main text does not show at all this symmetry the antidiagonal being of opposite sign compared to the diagonal. Our results thus confirm the theoretical prediction of [6] that the positive anti-diagonal disappears in the case of a finite correlation length. More precisely, it is shown in [6] that one expects the positive anti-diagonal to vanish as $L/l_\phi$ where $L$ is the system size. Moreover, for a system in the thermodynamic limit, a Luttinger Liquid approach shows that $G(k, k')$ is a regular function except for a singular diagonal bunching term. It is mainly negative, in particular around the anti-diagonal, for momenta $k \lesssim 1/l_\phi$.

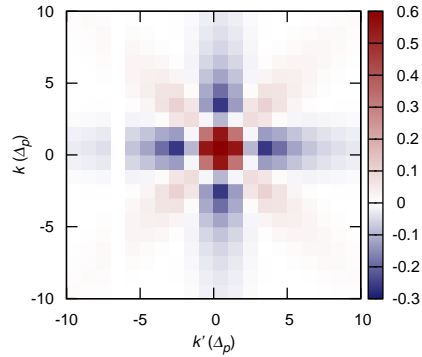

FIG. 4. Prediction for true long range order : $G(k, k')$ computed with the Bogoliubov approach in [8], for the temperature and peak linear density of Data C, convolved with the imaging response function.



\end{document}